\documentclass[fleqn,10pt]{wlscirep}
\usepackage[utf8]{inputenc}
\usepackage[T1]{fontenc}
\usepackage{upgreek}
\usepackage[capitalise]{cleveref}

\newcommand{\Wcmsqd}{{\mathrm{W}\mathrm{cm}^{-2}}}

\newcommand{\micron}{{\upmu\mathrm{m}}}
\newcommand{\ava}[1]{\textcolor{black}{#1}}
\newcommand{\tom}[1]{\textcolor{black}{#1}}

\newcommand{\rc}[1]{\textcolor{black}{#1}}
\newcommand{\bc}[1]{\textcolor{black}{#1}}
\newcommand{\mc}[1]{\textcolor{black}{#1}}

\title{Dominance of $\gamma$-$\gamma$ electron-positron pair creation\\in a plasma driven by high-intensity lasers}

\author[1]{Y. He}
\author[2]{T. G. Blackburn}
\author[3]{T. Toncian}
\author[1,*]{A. Arefiev}
\affil[1]{University of California at San Diego, La Jolla, California 92093, USA}
\affil[2]{Department of Physics, University of Gothenburg, SE-41296 Gothenburg, Sweden}
\affil[3]{Institute for Radiation Physics, Helmholtz-Zentrum Dresden-Rossendorf e.V., 01328 Dresden, Germany}

\affil[*]{aarefiev@eng.ucsd.edu}

\begin{abstract}
Creation of electrons and positrons from light alone is a basic prediction of quantum electrodynamics, but yet to be observed. Our simulations show that the required conditions are achievable using a high-intensity two-beam laser facility and an advanced target design. Dual laser irradiation of a structured target produces high-density $\gamma$ rays that then create ${>}10^8$ positrons at intensities of $2 \times 10^{22}~\Wcmsqd$. The unique feature of this setup is that the pair creation is primarily driven by the linear Breit-Wheeler process ($\gamma\gamma \to e^+ e^-$), which dominates over the nonlinear Breit-Wheeler and Bethe-Heitler processes. The favorable scaling with laser intensity of the linear process prompts reconsideration of its neglect in simulation studies and also permits positron jet formation at experimentally feasible intensities. Simulations show that the positrons, confined by a quasistatic plasma magnetic field, may be accelerated by the lasers to energies $> 200$~MeV.
\end{abstract}
\begin{document}

\flushbottom
\maketitle
\thispagestyle{empty}

\section*{Introduction}

High-power lasers, focused close to the diffraction limit, create ultrastrong electromagnetic fields that can be harnessed to drive high fluxes of energetic particles and to study fundamental physical phenomena~\cite{zhang.pop.2020}. At intensities exceeding $10^{23}~\Wcmsqd$, those energetic particles can drive nonlinear quantum-electrodynamical (QED) processes~\cite{erber.rmp.1966,dipiazza.rmp.2012} otherwise only found in extreme astrophysical environments~\cite{harding.rpp.2006,ruffini.pr.2010}. One such process is the creation of electron-positron pairs from light alone. Whereas multiphoton (nonlinear) pair creation has been measured once, using an intense laser~\cite{burke.prl.1997}, the two-photon process ($\gamma\gamma \to e^+ e^-$, referred to here as the linear Breit-Wheeler process~\cite{breit.pr.1934}) has yet to be observed in the laboratory with real photons. As the probability of the nonlinear process grows nonperturbatively with increasing field strength~\cite{reiss.jmp.1962,ritus.jslr.1985}, it is expected to provide the dominant contribution to pair cascades in high-field environments, including laser-matter interactions beyond the current intensity frontier~\cite{bell.prl.2008,ridgers.prl.2012} and pulsar magnetospheres~\cite{timokhin.apj.2019}.

The small size of the linear Breit-Wheeler cross section means that high photon flux is necessary for its observation. Achieving the necessary flux requires specialized experimental configurations~\cite{pike.np.2014,ribeyre.pre.2016} and therefore its possible contribution to \emph{in situ} electron-positron pair creation has hitherto been neglected in studies of high-intensity laser-matter interactions. However, these \ava{interactions} create not only regions of ultrastrong electromagnetic field, but also high fluxes of accelerated particles, because relativistic effects mean that even a solid-density target can become transparent to intense laser light~\cite{kaw.physfluids.1970,palaniyappan.natphys.2012}. In the situation of multiple colliding laser pulses, which is the most advantageous geometry for driving nonlinear QED cascades~\cite{bell.prl.2008,zhu.ncommun.2016,grismayer.pre.2017,gonoskov.prx.2017}, there are, as a consequence, dense, counterpropagating flashes of $\gamma$ rays, and so the neglect of linear pair creation may not be appropriate.

\ava{Recent construction of multi-beam high-intensity laser facilities, such as ELI-Beamlines~\cite{weber.mre.2017}, ELI-Nuclear Physics~\cite{gales.rpp.2018,lureau.hplse.2020}, and Apollon~\cite{papadopoulos.hpl.2016}, and a significant progress in fabrication of $\micron$-scale structured targets~\cite{snyder.pop.2019,bailly-grandvaux.pre.2020} open up qualitatively novel regimes of pair production for exploration. Specifically, we show that a structured plasma target irradiated by two laser beams creates an environment where the linear process dominates over the nonlinear and over the Bethe-Heitler process. Remarkably, this regime does not require laser intensities beyond than what is currently available. At $I_0 < 5 \times 10^{22}~\Wcmsqd$, the positron yield from the linear process is $\sim 10^9$, which is four orders of magnitude greater than that envisaged in Refs.~[\citeonline{pike.np.2014}] and [\citeonline{ribeyre.pre.2016}]}. These positrons are generated when two high-energy electron beams, accelerated by and copropagating with laser pulses that are guided along a plasma channel, collide head-on, emitting synchrotron photons that collide with each other and the respective oncoming laser. Not only does this provide an opportunity to study the linear Breit-Wheeler process itself, which is of interest because of its role in astrophysics~\cite{gould.prl.1966,bonometto.mnras.1971,piran.rmp.2005}, but also the transition between linear and nonlinear-dominated pair cascades. In an astrophysical context, the balance between these two determines how a pulsar magnetosphere is filled with plasma; as in the laser-plasma scenario, the controlling factors are the field strength and photon flux~\cite{burns.apj.1984,zhang.aa.1998,voisin.mnras.2018,chen.apj.2020}. We also show that the positrons, created inside the plasma channel coterminously with the laser pulses, may be confined and accelerated to energies of hundreds of MeV, which raises the possibility of generating positron jets. \bc{The transverse confinement needed to accelerate positrons is provided by a slowly evolving plasma magnetic field.
Crucially, it is the same field that enables acceleration of the ultra-relativistic electrons prior to the collision of the two laser pulses.}


    \begin{figure}
    \centering
    \includegraphics[width=0.5\linewidth]{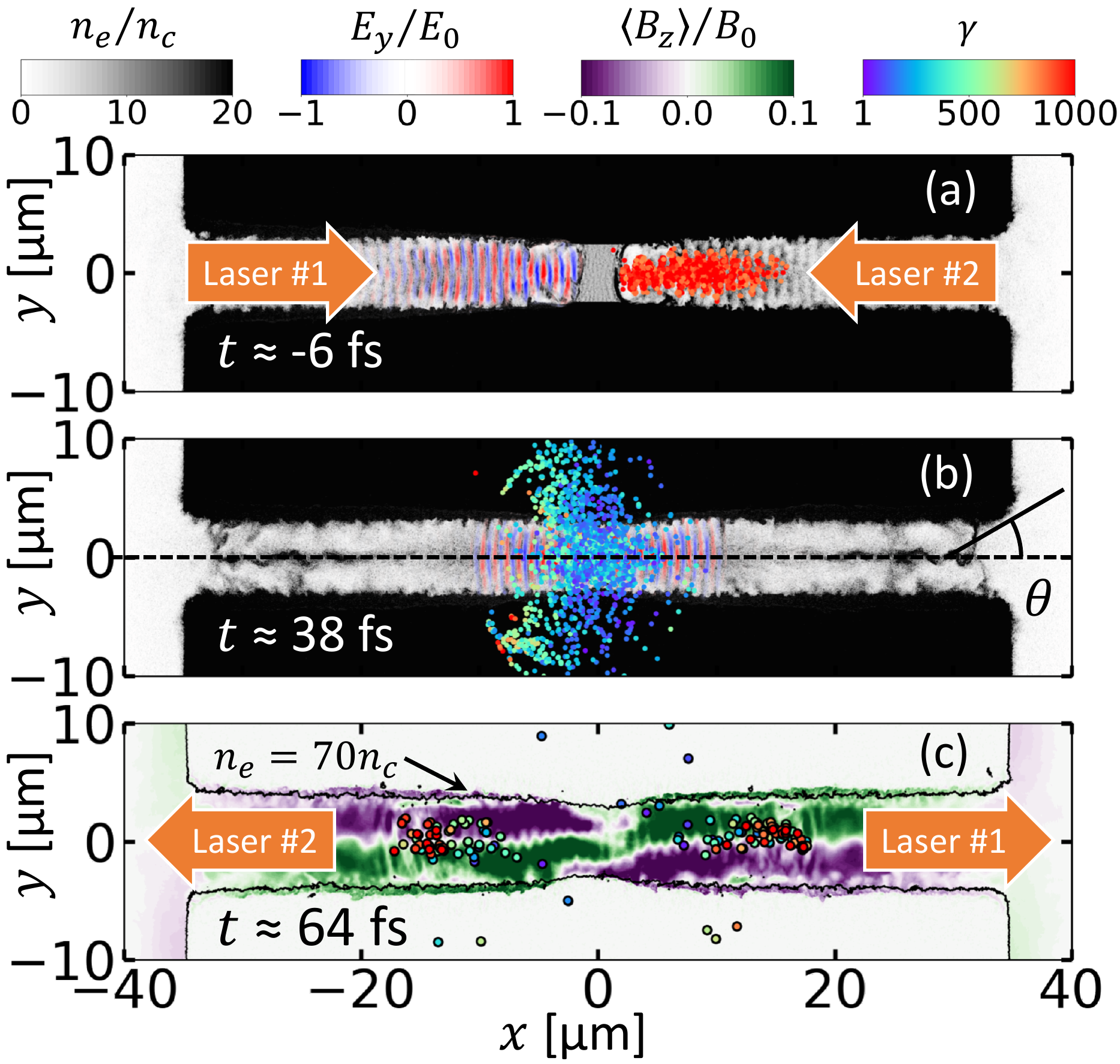}
    \caption{%
        Structured plasma target during irradiation by two laser pulses ($a_0 = 190$) \bc{in a 2D PIC simulation}.
        (a) Electron density (gray scale), transverse electric field of laser \#1 (color scale) and energetic electrons with $\gamma \geq 800$ accelerated by laser \#2 (dots, colored by $\gamma$).
        (b) Total transverse electric field (color scale) and electrons from panel (a).
        (c) Laser-accelerated positrons (points), confined by the quasistatic plasma magnetic field $\langle B_z \rangle$ (color scale).
        $E_0$ and $B_0$ are the peak laser electric and magnetic fields in vacuum.}
    \label{fig:schematic}
    \end{figure}

\section*{Results}

The target configuration considered in this work is shown in \cref{fig:schematic}(a). A structured plastic target with a pre-filled channel is irradiated from both sides by two 50-fs, high-intensity laser pulses that have the same peak normalized laser amplitude $a_0$, in the range $100 \leq a_0 \leq 190$. Here $a_0 = 0.85  I_0^{1/2} [10^{18} \Wcmsqd] \lambda_0[\micron]$, where $I_0$ is the peak intensity of the laser and $\lambda_0 = 1~\micron$ its wavelength in vacuum. The target structure, where a channel of width $d_\text{ch} = 5~\micron$ and electron density $n_e = (a_0/100) 3.8 n_c$ is embedded in a bulk with higher density $n_e = 100 n_c$, enables stable propagation~\cite{stark.prl.2016} and alignment of the two lasers. Here $n_c = \pi m c^2 / (e \lambda_0)^2$ is the so-called critical density, where $e$ is the elementary charge, $m$ is the electron mass, and $c$ is the speed of light. \bc{At relativistic laser intensities ($a_0 \gg 1$), the cutoff density for the laser increases roughly linearly with $a_0$ due to relativistically induced transparency. Scaling the channel density with $a_0$ ensures that the optical properties of the channel and thus the phase velocity of the laser wave-fronts are approximately unchanged with increase of $a_0$.} Structured targets with empty channels have successfully been used in experiments~\cite{snyder.pop.2019,bailly-grandvaux.pre.2020} and it is now possible to fabricate targets with prefilled channels, similar to those considered in this work~\footnote{J. Williams, Private communications at General Atomics, 2019.}.

    \begin{figure}[htb]
    \centering
    \includegraphics[width=0.95\linewidth]{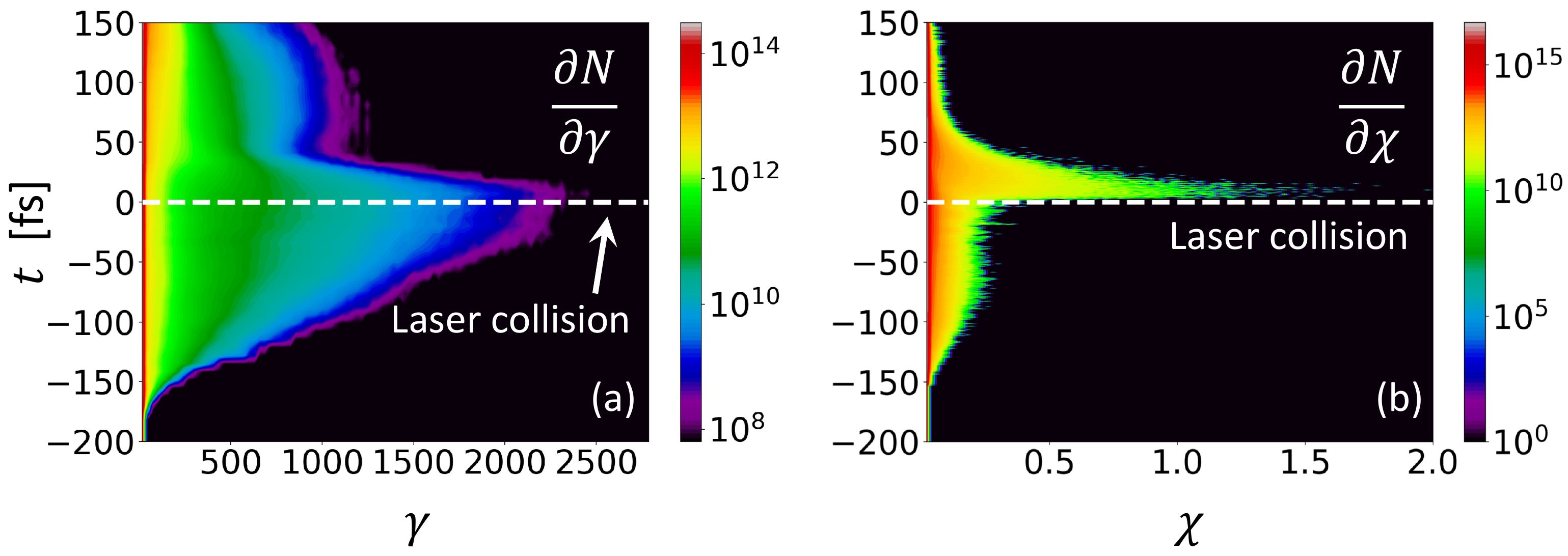}
    \caption{\bc{
    Time evolution of the distributions of electron (left panel) relativistic factor $\gamma$ and (right panel) quantum nonlinearity parameter $\chi$ (defined by \cref{Eq:chi}) for the 2D PIC simulation shown in \cref{fig:schematic}. The two laser pulses have $a_0 = 190$ and collide at $t = 0$.}}
    \label{fig:energy and eta}
    \end{figure}

    \begin{figure} [htb]
    \centering
    \includegraphics[width=0.99\linewidth]{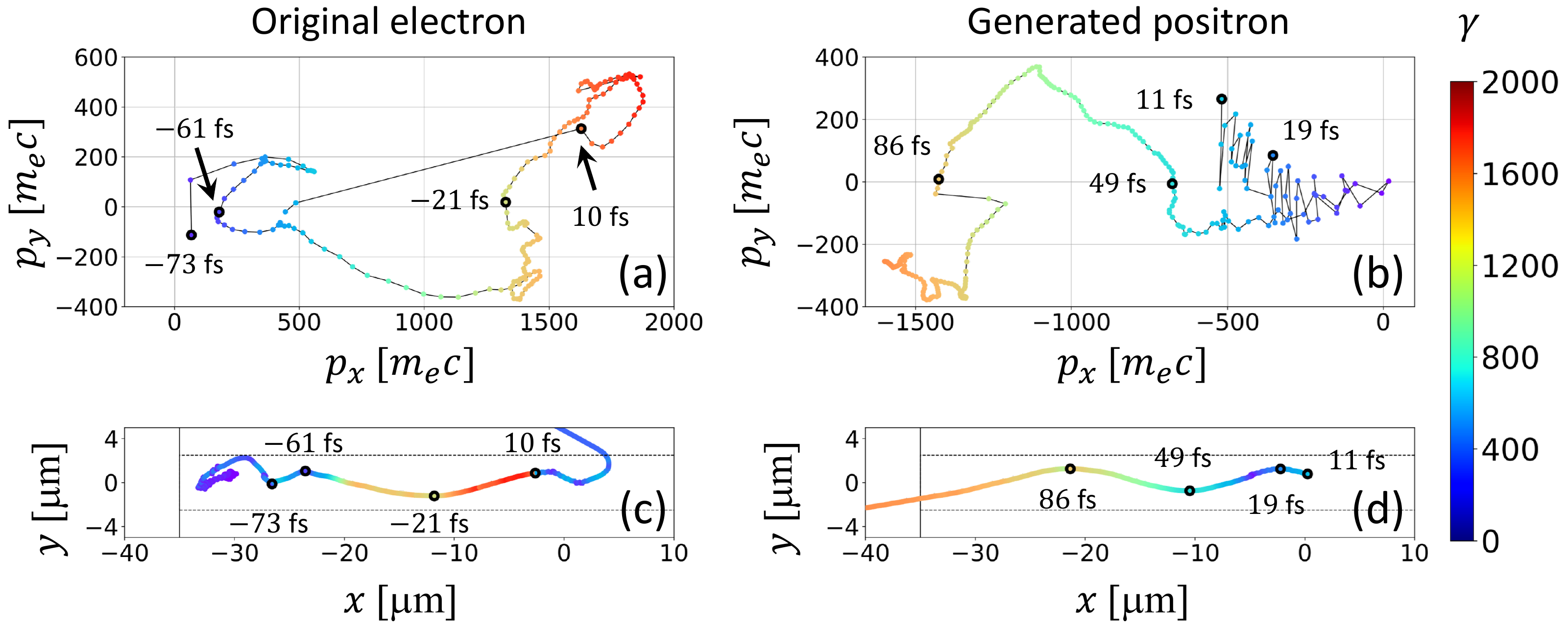}
    \caption{\bc{Trajectories of an accelerated plasma electron and a generated positron from the 2D PIC simulation shown in \cref{fig:schematic}:
    (a, b) tranverse momenta $p_x, p_y$ and (c, d) position in the $x$-$y$ plane.
    Color coding denotes the magnitude of the relativistic factor $\gamma$.
    The vertical solid line is the initial position of the left edge of the target. The horizontal dashed lines show the initial location of the channel walls.
    The timestep between the colored markers is 0.5~fs.
    Timestamps are provided for selected markers (shown as dark circles) to facilitate comparison between trajectories in $(p_x,p_y)$-space and $(x,y)$-space.
    To improve visibility, the electron trajectory in (a) is shown for $-73 \mbox{ fs} \leq t \leq 11$~fs.}}
    \label{fig:e and p trajectories}
    \end{figure}

The interaction is simulated in 2D with the fully relativistic particle-in-cell (PIC) code EPOCH~\cite{arber2015epoch}, which includes Monte Carlo modules for quantum synchrotron radiation and nonlinear pair creation~\cite{ridgers.jcp.2014}. \mc{At each time-step, the quantum synchrotron radiation module computes the quantum nonlinearity parameter,
\begin{equation} \label{Eq:chi}
    \chi \equiv \frac{\gamma}{E_S} \sqrt{\bigg(\bf{E} + {\rm{\frac{1}{c}}}[v \times B]\bigg){\rm{^2}} - {\rm{\frac{1}{c^2}}}(E\cdot v){\rm{^2}}},
\end{equation}
for each charged macro-particle using the electric and magnetic fields ($\bf{E}$ and $\bf{B}$) at the particle location, as well as the particle relativistic factor $\gamma$ and velocity $\bf{v}$.
Here $E_S \approx 1.3 \times 10^{18} \ \rm{V/m}$ is the Schwinger field~\cite{sauter.zp.1931,heisenberg.zp.1936,schwinger.pr.1951}.
The parameter $\chi$ controls the total radiation power and the energy spectrum of the emitted photons.
In the quantum regime $\chi \gtrsim 1$, which is reached in this work, it is necessary to take into account the recoil experienced by the particle when emitting individual photons.
This is done self-consistently by the PIC simulation, which uses the Monte Carlo algorithm described in Refs.~[\citeonline{ridgers.jcp.2014,gonoskov.pre.2015}].
} Note that, since the ion species is fully ionized carbon, Bethe-Heitler pair creation, already demonstrated in laser-driven experiments~\cite{chen.prl.2009,sarri.prl.2013}, may be neglected. Detailed simulation and target parameters are provided in \ava{the Methods section}. All the results presented in this \ava{paper} have been appropriately normalized by taking the size of the ignored dimension to be equal to the channel width $d_\text{ch}$, i.e. $5~\micron$.

The plasma channel, being relativistically transparent to the intense laser light~\cite{kaw.physfluids.1970,palaniyappan.natphys.2012}, acts as an optical waveguide. The laser pulses propagate with nearly constant transverse size through the channel, pushing plasma electrons forward. This longitudinal current generates a slowly evolving, azimuthal magnetic field with peak magnitude $0.6$~MT (30\% of the laser magnetic field strength) at $a_0 = 190$, as shown in \cref{fig:schematic}(c). The magnetic field enables confinement and direct laser acceleration of the electrons~\cite{stark.prl.2016,gong.PhysRevE.2020}. After propagating for $\sim$30~$\micron$ along the channel, laser \#2 in \cref{fig:schematic}(a) has accelerated a left-moving, high-energy, high-charge electron beam that performs transverse oscillations of amplitude $\sim$2~$\micron$: the number of electrons with relativistic factor $\gamma > 800$ is $4 \times 10^{11}$, which is equivalent to a charge of 64~nC. Laser~\#1 generates a similar population of electrons moving to the right, \bc{with a representative electron trajectory shown in \cref{fig:e and p trajectories}(c).}

\bc{The plasma magnetic field has an essential role in enabling generation of ultrarelativistic electrons.
Transverse deflections by the magnetic field keep $p_y$ antiparallel to the transverse electric field $E_y$ of the laser, despite the oscillation of the latter. As a result, the electron continues to gain energy while moving along the channel and performing transverse oscillations, as may be seen in \cref{fig:e and p trajectories}(a) and \cref{fig:e and p trajectories}(c). In the absence of the magnetic field, the oscillations of $E_y$ would terminate the energy gain prematurely. The magnetic field of the plasma has to be sufficiently strong to ensure that the electron deflections occur on the same time scale as the oscillations of $E_y$. This criterion can be formulated in terms of the longitudinal plasma current and is provided in Ref.~[\citeonline{gong.PhysRevE.2020}]. The time evolution of the electron spectrum, which shows this mechanism in action, is shown in \cref{fig:energy and eta}(a).}

    \begin{figure}[htb]
    \centering
    \includegraphics[width=0.99\linewidth]{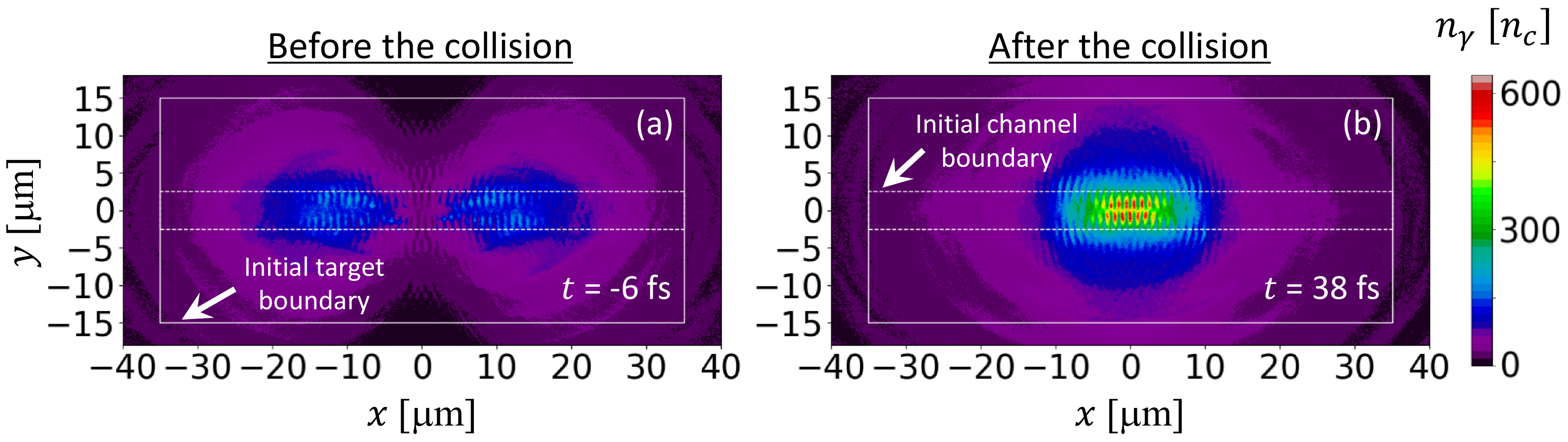}
    \caption{\rc{The density of photons with energy $\varepsilon_{\gamma} \geq 1$~keV before (a) and after (b) the collision of two laser pulses ($a_0 = 190$) inside the structured target. Data from the 2D PIC simulation shown in \cref{fig:schematic}.}}
    \label{fig:photon density}
    \end{figure}

The target length is such that no appreciable depletion of the laser pulses occurs by the time they reach the midplane ($x = 0$), $t = 0$. Here, the high-energy electron beams collide head-on with the respective oncoming laser pulse, each of which
\rc{has an intensity at least as large as its initial value (the magnitude can increase slightly due to pulse shaping during propagation along the channel)}.
\bc{This configuration maximizes the quantum nonlinearity parameter $\chi$ for the electrons, as the two terms under the square root in \cref{Eq:chi} are additive for counterpropagation.
In copropagation, by contrast, they almost cancel each other.
This is why radiation prior to the collision, when electrons propagate in the same direction as the accelerating laser pulse, is driven primarily by the plasma magnetic field.
As is shown in \cref{fig:energy and eta}(b), the largest value of $\chi \approx 0.25$ until the collision occurs; immediately thereafter, the cancellation is eliminated and $\chi$ increases rapidly to approximately $1.25$.
}
\Cref{fig:schematic}(b) shows the impact of the collision on the energetic electrons from \cref{fig:schematic}(a): they radiate away a substantial fraction of the energy they gained during the acceleration phase and are scattered out of the channel. \bc{Similar behavior is shown in \cref{fig:e and p trajectories}(c): the electron encounters the counterpropagating laser beam at about $t = 10$~fs and then its energy decreases rapidly.}

    \begin{figure}[htb]
    \centering
    \includegraphics[width=0.45\linewidth]{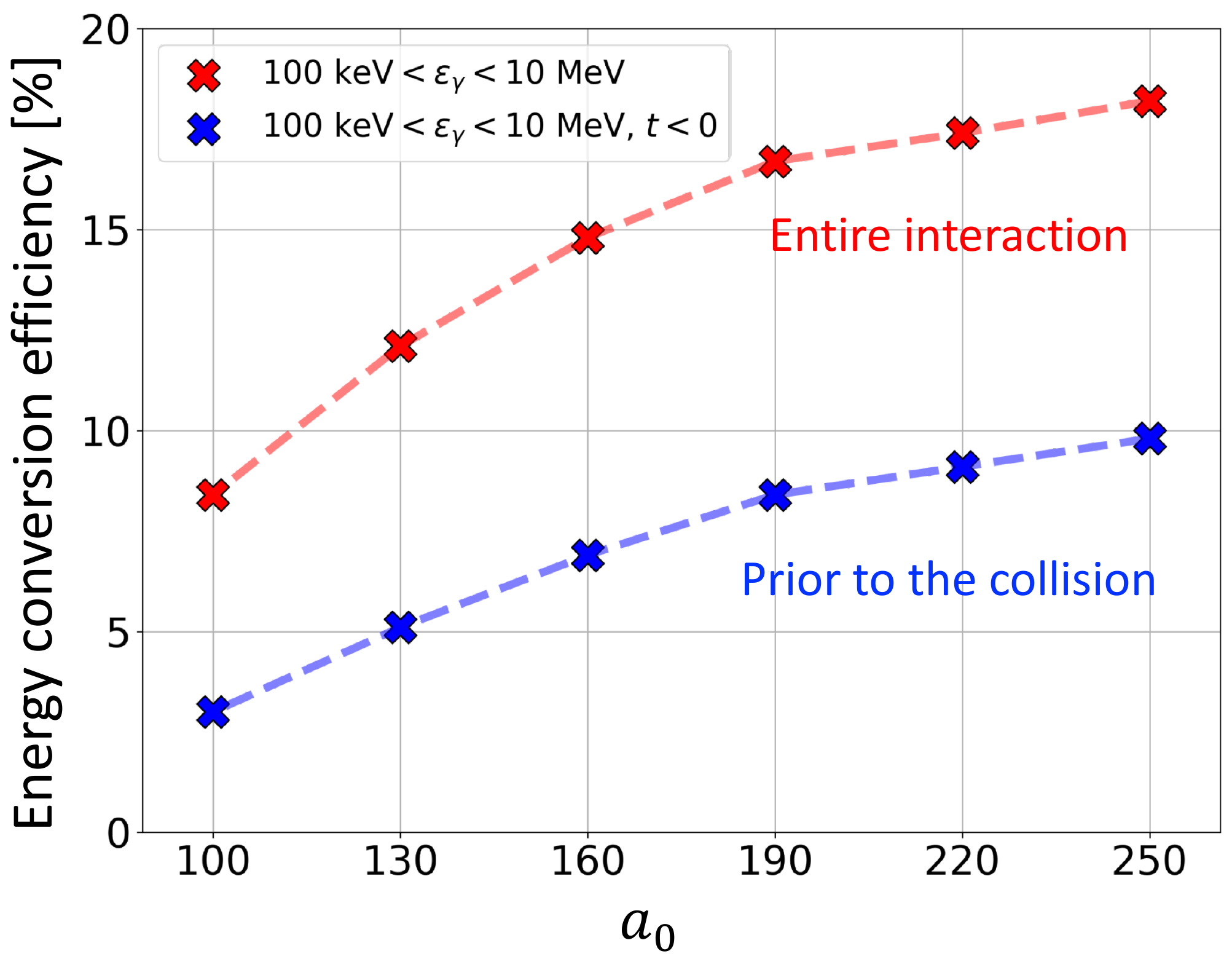}
    \caption{\rc{Conversion efficiency of the laser energy into $\gamma$ rays with energies $100~\mbox{keV} \leq \varepsilon_{\gamma} \leq 10$~MeV, in 2D PIC simulations with two counterpropagating laser pulses that have the same peak normalized laser amplitude $a_0$. The blue markers show the conversion efficiency recorded \emph{before} the two pulses collide, whereas the red markers show the conversion efficiency computed over the duration of the entire laser-target interaction.}}
    \label{fig:conversion efficiency}
    \end{figure}

\rc{
The observed increase in $\chi$ during the electron-laser collision increases the radiation power of the individual electrons.
This emission occurs in a highly localized region, which leads to the marked increase in photon density shown in \cref{fig:photon density} for the case where $a_0 = 190$. The conversion efficiency of the laser energy into photons with energies $100~\mbox{keV} \leq \varepsilon_{\gamma} \leq 10$~MeV is shown in \cref{fig:conversion efficiency} over a wide range of $a_0$. We are interested in the photons in this energy range because these are the photons that participate in the linear Breit-Wheeler process in our setup, as shown in \cref{fig:energy vs energy} of the Methods section. As expected, there is a significant increase in the conversion rate caused by the electron-laser collision.}
The configuration under consideration here therefore represents a micron-scale, plasma-based realization of an all-optical laser--electron-beam collision (see also Ref.~[\citeonline{zhu.ncommun.2016}]). This geometry is the subject of theoretical~\cite{neitz.prl.2013,blackburn.prl.2014,vranic.njp.2016} and experimental~\cite{cole.prx.2018,poder.prx.2018} investigation into radiative energy loss in the quantum regime, as well as nonlinear pair creation~\cite{lobet.prab.2017}. \ava{It is worth emphasizing that the use of the structured target has two key benefits compared to the commonly used gas targets: automatic alignment of the colliding electrons with an oncoming laser beam and \bc{a considerably higher density of colliding electrons.}}

\begin{figure}[htb]
    \centering
    \includegraphics[width=0.5\linewidth]{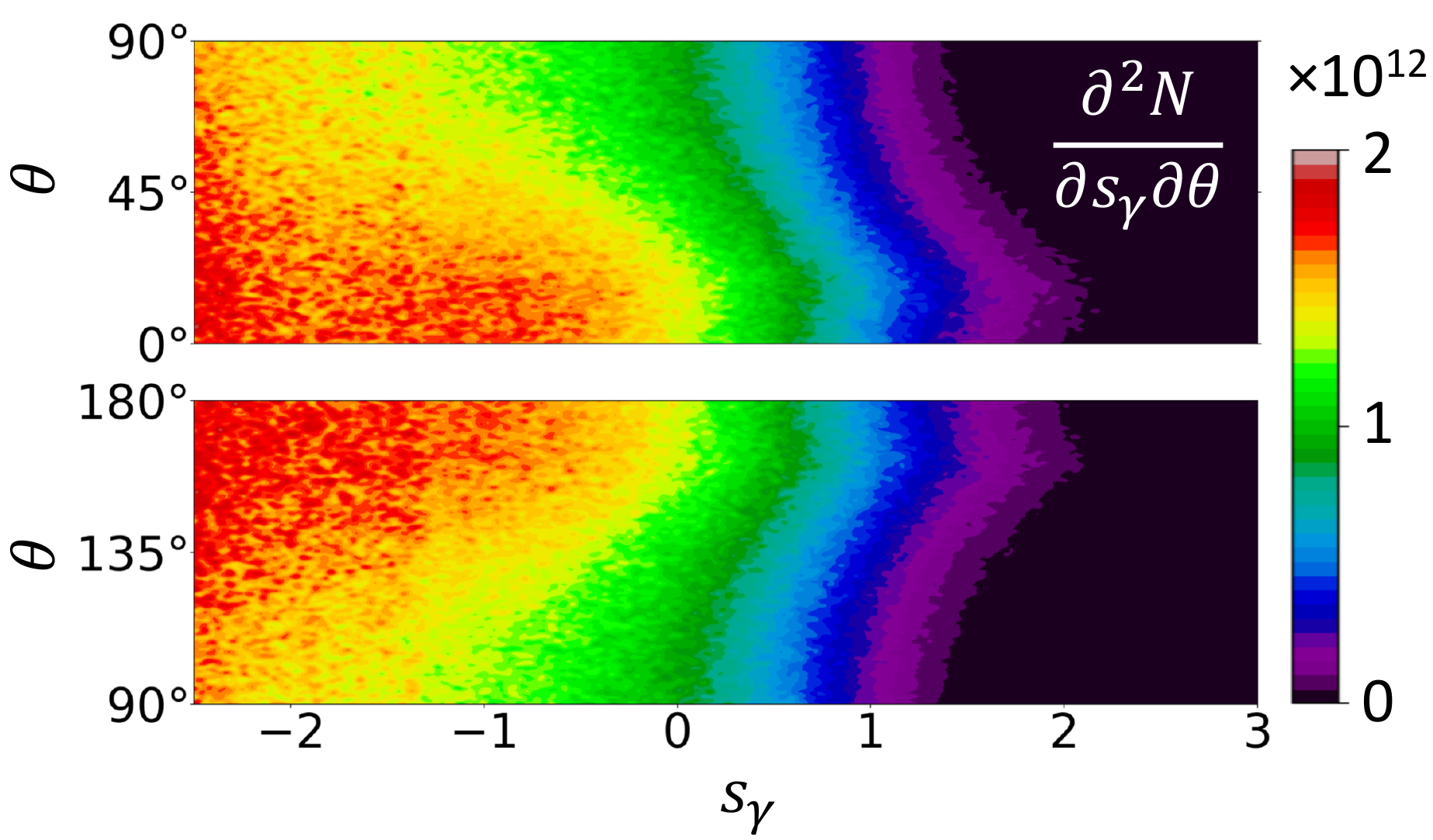}
    \caption{%
        Energy-angle spectrum, $\partial ^2 N/(\partial s_{\gamma}\partial\theta)$ [$^{{\circ}{-1}}$], of the photons emitted inside the channel \bc{in the 2D PIC simulation from Fig.~\ref{fig:schematic}}, where $\theta$ is the angle defined in Fig.~\ref{fig:schematic}(b) and $s_\gamma \equiv \log_{10} (\epsilon_\gamma [\text{MeV}])$.
        The spectrum for $-180^{\circ} \leq \theta \leq 0^{\circ}$ is similar.}
    \label{fig:photon_distribution}
    \end{figure}

The angularly resolved spectrum of the emitted photons is shown in \cref{fig:photon_distribution}. There are approximately $2\times10^{14}$ photons with energies between 100~keV and 10~MeV and with $90^{\circ} \leq \theta \leq 180^{\circ}$. This is essentially half of the energetic photons emitted by the left-moving electrons (the other half is emitted with $-180^{\circ} \leq \theta \leq -90^{\circ}$ and has a similar spectrum). The photons emitted by one electron beam collide with both the oncoming laser and the photons emitted by the other electron beam. The former drives electron-positron pair creation by the nonlinear Breit-Wheeler process, $\gamma \xrightarrow{\text{EM field}} e^+ e^-$~\cite{erber.rmp.1966,ritus.jslr.1985}: at $a_0 = 190$, our simulations predict a yield of $5 \times 10^8$ pairs.

\begin{figure}
    \centering
    \includegraphics[width=0.9\columnwidth]{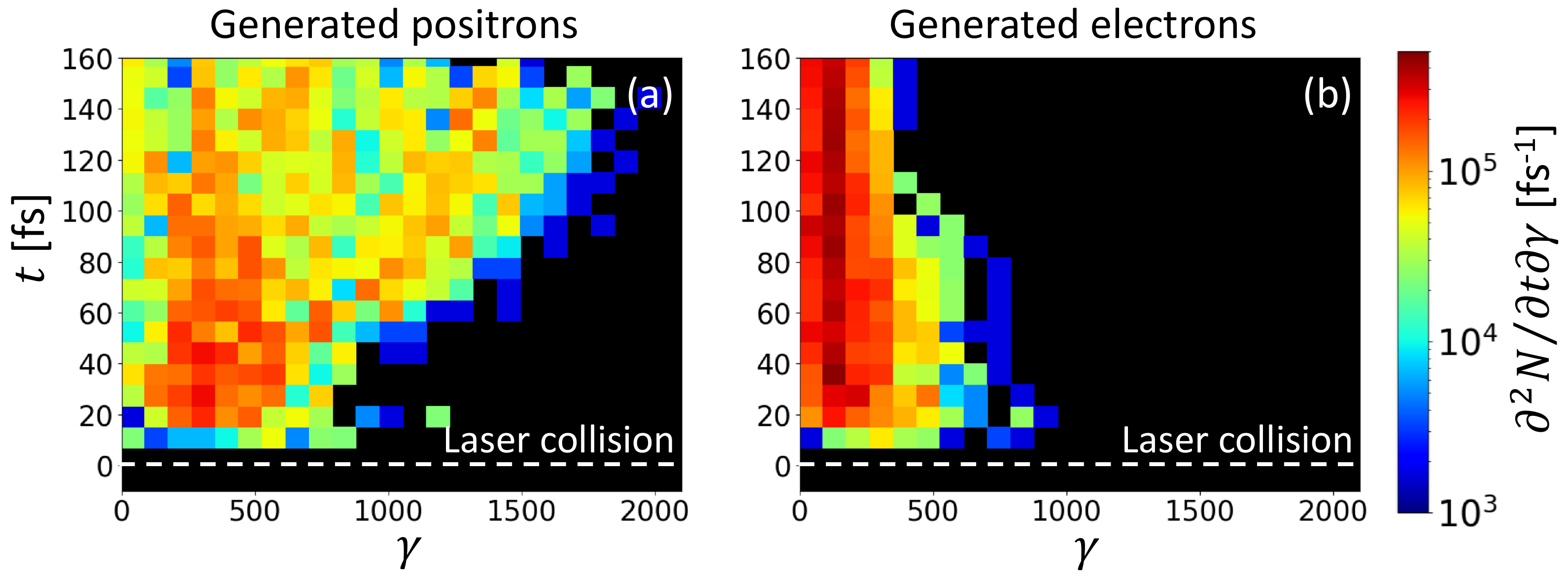}
    \caption{\bc{Time evolution of the energy spectra of electrons and positrons generated by nonlinear Breit-Wheeler pair creation. Data from the 2D PIC simulation shown in \cref{fig:schematic} at $a_0 = 190$.}}
    \label{fig:produced spectra}
    \end{figure}

The positrons subsequently undergo direct laser acceleration in much the way as the electrons: PIC simulations show that the typical relativistic factor of a right-moving positron increases to $\gamma \approx 1000$ as it propagates from $x \approx 0$ to $x \approx 20~\micron$.
\bc{This is illustrated in \cref{fig:schematic}(c) and corroborated by the time evolution of the positron energy spectrum shown in \cref{fig:produced spectra}(a). A representative trajectory for a positron moving from the central region towards the left target boundary is shown in \cref{fig:e and p trajectories}(d). Acceleration is made possible by the plasma magnetic field, which is confining (on the left-hand side of the target) for electrons moving to the right, or equivalently, positrons moving to the left [compare \cref{fig:e and p trajectories}(c) and \cref{fig:e and p trajectories}(d)].} Crucially, \cref{fig:schematic}(c) shows that this magnetic field \bc{polarity} is preserved well after the lasers and electron beams collide. \bc{This is why, after the two laser pulses collide and pass through each other, they can accelerate the positrons, but not the electrons, created in by photon-photon collisions, as seen in \cref{fig:produced spectra}. The generated electrons are not transversely confined in our magnetic field configuration when moving from the center towards either of the channel openings.} However, the continued propagation of the lasers along the channel raises the possibility of accelerating positron jets, if there is sufficient pair creation in the channel center.

We now show that this is the case, and furthermore that the pair creation is dominated by the \emph{linear} Breit-Wheeler process. The cross section is~\cite{breit.pr.1934}:
    \begin{equation}
    \sigma_{\gamma\gamma} = \frac{\pi r_e^2}{2 \varsigma}
        \left[
            (3 - \beta^4) \ln\!\left( \frac{1+\beta}{1-\beta} \right)
            -2 \beta (2 - \beta^2)
        \right],
    \label{eq:BWCrossSection}
    \end{equation}
where $r_e = e^2 / (m c^2)$ is the classical electron radius, $\beta = \sqrt{1 - 1/\varsigma}$, and $\sqrt{\varsigma}$ is the normalized center-of-mass energy, $\varsigma = \epsilon_1 \epsilon_2 (1 - \cos\psi) / (2 m^2 c^4)$, for two photons with energy $\epsilon_{1,2}$ colliding at angle $\psi$. \Cref{eq:BWCrossSection} is the cross section for two-photon pair creation in vacuum: while it is modified by a strong EM field~\cite{ng.prd.1977,kozlenkov.jetp.1986,hartin.ijmpa.2018,hartin.2006}, these corrections, which scale as $(\chi_\gamma / \varsigma)^2$ for photon quantum nonlinearity parameter $\chi_\gamma$~\cite{bks}, are negligible for the scenario under consideration here (see the Supplemental Material for details).

We take as a representative value $\sigma_{\gamma\gamma} \approx 2 r_e^2$ (approximately its maximum, at $\varsigma \approx 2$) and assume that we have two photon populations of number density $n_\gamma$, colliding head-on in a volume of length $c \tau$ (the laser pulse length) and width $d_\text{ch}$ (the width of the channel).
The number of photons (in each beam) is $N_\gamma \approx 10^9 \lambda_0[\micron] P_\gamma (n_e / n_c) (c \tau / \lambda_0) (d_\text{ch} / \lambda_0)^2$, where $P_\gamma$ is the number of photons emitted per electron, $n_e$ is the electron number density and $\lambda_0$ is the laser wavelength.
The number of positrons produced, $N_\text{lin}^\text{BW} = 2 N_\gamma^2 \sigma_{\gamma\gamma} / d_\text{ch}^2$, follows as $N_\text{lin}^\text{BW} \approx 40 P_\gamma^2 (n_e/n_c)^2 (c \tau / \lambda_0)^2 (d_\text{ch} / \lambda_0)^2$.
The physical parameters are $n_e = 7 n_c$, $\tau = 50$~fs, $d_\text{ch} = 5~\micron$, and $\lambda_0 = 1~\micron$.
The number of photons emitted per laser period by a counterpropagating electron is $P_\gamma \approx 18 \alpha a_0$, where $\alpha \simeq 1/137$ is the fine-structure constant.
By setting $P_\gamma = 20$, we obtain a total number of photons, $2 N_\gamma \approx 1.3 \times 10^{14}$, which is approximately consistent with the simulation result.
As a consequence, we predict that $N_\text{lin}^\text{BW} \approx 7 \times 10^9$.
Given that $P_\gamma \propto a_0$ and $n_e \propto a_0$, we predict a scaling of $N_\text{lin}^\text{BW} \propto a_0^4$.

    \begin{figure}[htb]
    \centering
    \includegraphics[width=0.6\columnwidth]{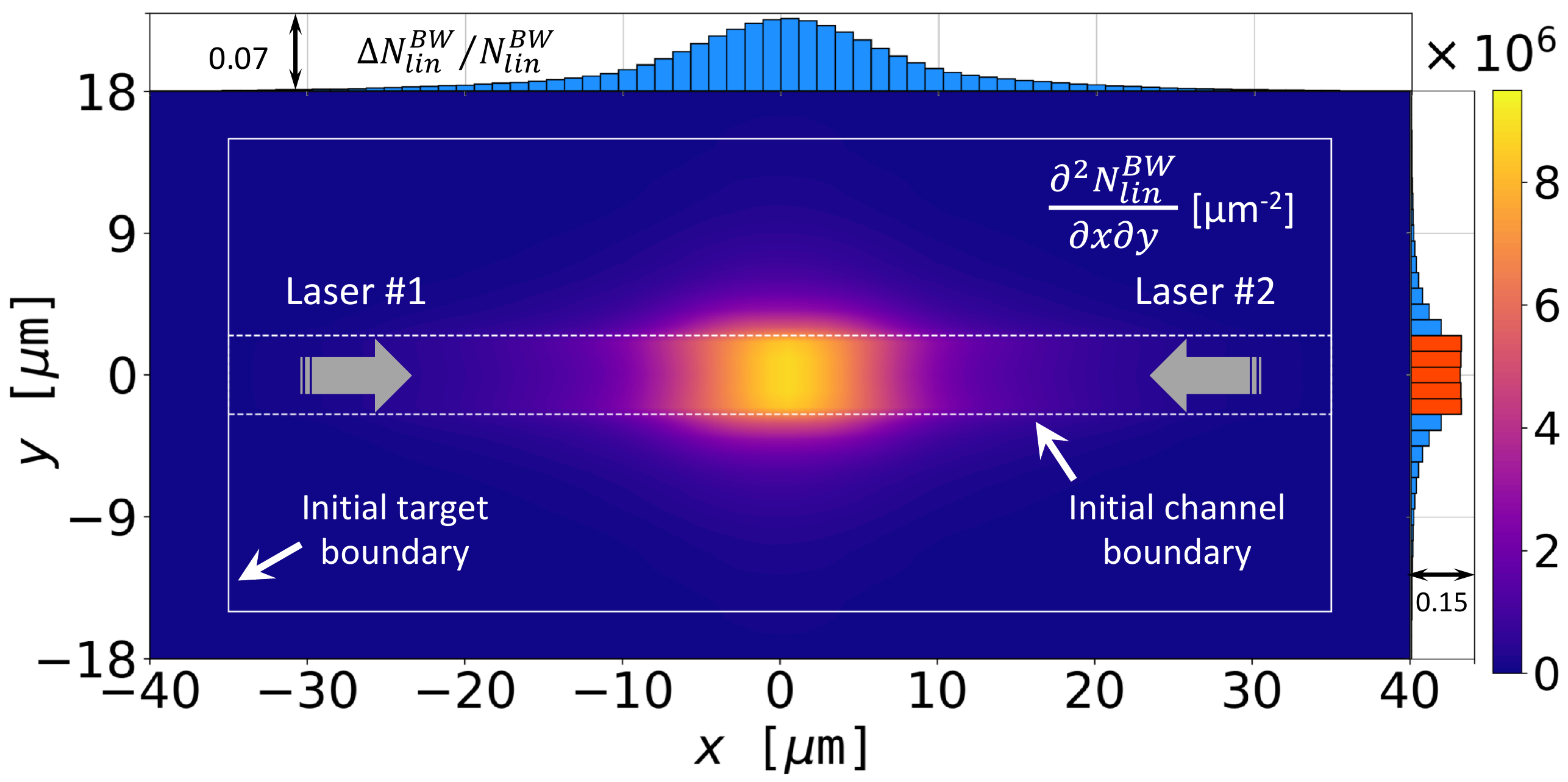}
    \caption{%
        (color scale) Probability density that an electron-positron pair is created by the linear Breit-Wheeler process at longitudinal and transverse coordinate $x$ and $y$, when a structured target is irradiated by two laser pulses with peak normalized amplitude $a_0 = 190$. \bc{The result is obtained by post-processing the 2D PIC simulation from \cref{fig:schematic} using the algorithm described in Methods.} The density integrated over $x$ ($y$), and normalized to the total number of pairs, is shown to the right (above).
        }
    \label{fig:position}
    \end{figure}

This is considerably larger than the number of pairs expected from the nonlinear Breit-Wheeler process; moreover, as the probability rate for the latter is exponentially suppressed with decreasing $a_0$, we expect the yield to be much more sensitive to reductions in laser intensity.
The potential dominance of the linear process motivates a precise computation, which takes into the account the energy, angle and temporal dependence of the photon emission.

However, direct implementation of the linear Breit-Wheeler process in a PIC code is a significant computational challenge, as it involves binary collisions of macroparticles and the interaction must be simulated in at least 2D. The simulation at $a_0 = 190$ generates ${\sim}10^8$ macrophotons in the energy range relevant for linear Breit-Wheeler pair creation and therefore ${\sim}10^{16}$ possible pairings. This can be reduced by using bounding volume hierarchies~\cite{jansen.jcp.2018}, which is effective if the photon emission and the pair creation are well-separated in time and space.
In our case, there is no such separation.
As such, we postprocess the simulation output to obtain the yield of linear Breit-Wheeler pairs, using the algorithm described in Methods. Note that the photons used to compute this yield are the same photons used by the simulation to compute the yield of \emph{nonlinear} Breit-Wheeler pairs. As such, while the photon number would change if the simulation were performed in 3D rather than 2D, the yield of both processes would be affected in a similar way.

    \begin{figure}[htb]
    \centering
    \includegraphics[width=0.4\linewidth]{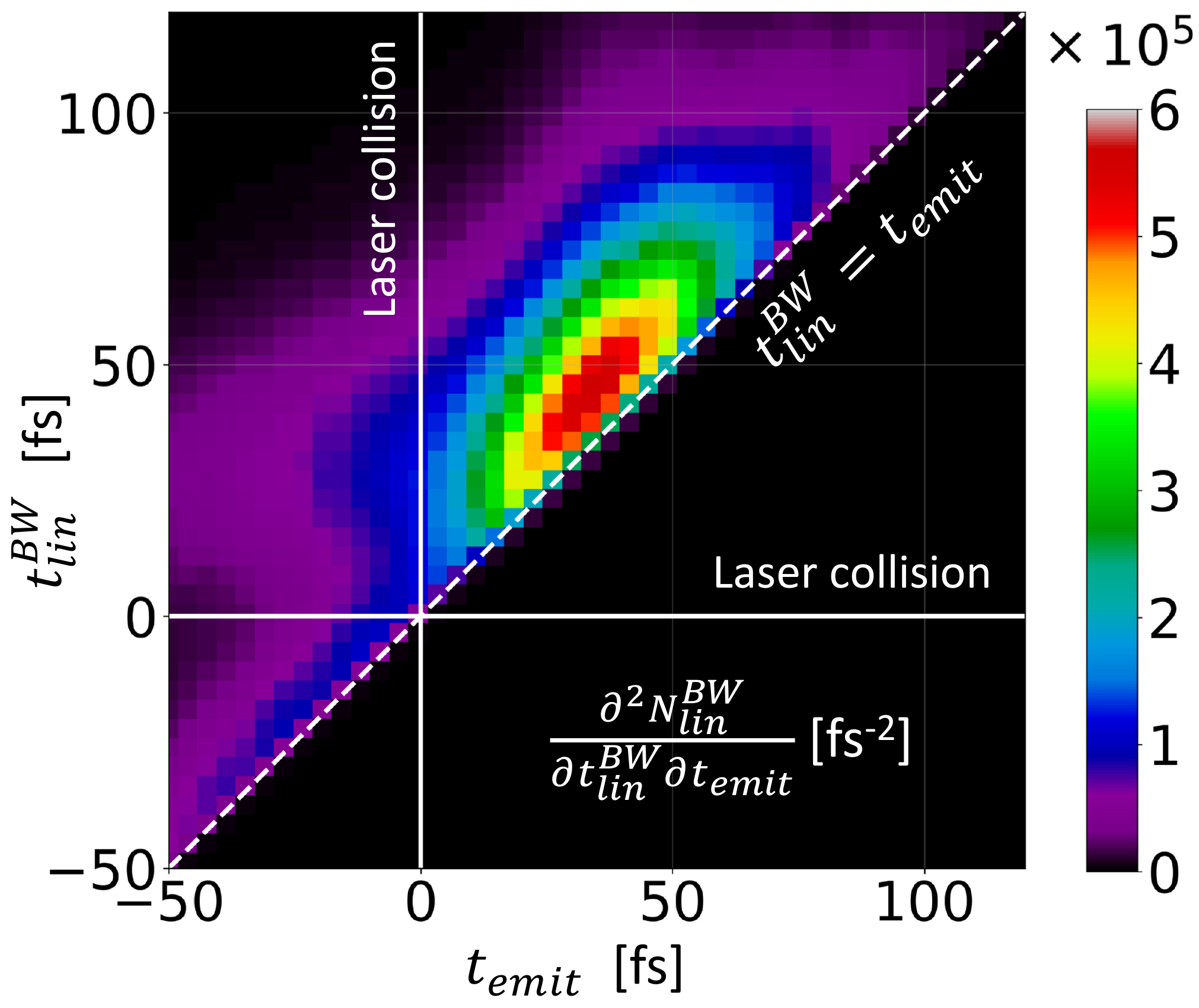}
    \caption{%
        (color scale) Probability density that an electron-positron pair is created by the linear Breit-Wheeler process at time $t_\text{lin}^\text{BW}$, by photons that were emitted at times $t_\text{emit}$, when a structured target is irradiated by two laser pulses with peak normalized amplitude $a_0 = 190$.
        The density includes a normalizing factor of $1/2$ because each pair has two parent photons. \bc{This plot is obtained by post-processing the 2D PIC simulation from \cref{fig:schematic} using the algorithm described in Methods.}
    }
    \label{fig:time_time_linear}
    \end{figure}

The location and time that pairs are created by the linear process, as determined by this algorithm for the case that $a_0 = 190$, are shown in \cref{fig:position} and \cref{fig:time_time_linear} respectively. Approximately 59\% of the pairs are created inside the original channel boundary. The majority (74\%) of pairs are created by photons emitted after $t = 0$, when the high-energy electrons collide with the respective counterpropagating laser. There is a smaller contribution from photons that are emitted during the acceleration phase, $t_\text{emit} < 0$; radiation in this case is driven by the plasma magnetic field, because the energetic electrons are moving in the same direction as the laser~\cite{stark.prl.2016,wang.PhysRevApplied.2020}. The dominance of the post-collision contribution is caused by the increase in the quantum parameter $\chi_e$ for counterpropagation. The fact the pair creation overlaps with the laser pulses (in both time and space) indicates that the positrons could be accelerated out of the channel, as the magnetic field, shown in \cref{fig:schematic}(c), has the correct orientation to confine them.

\begin{figure}
    \centering
    \includegraphics[width=0.6\linewidth]{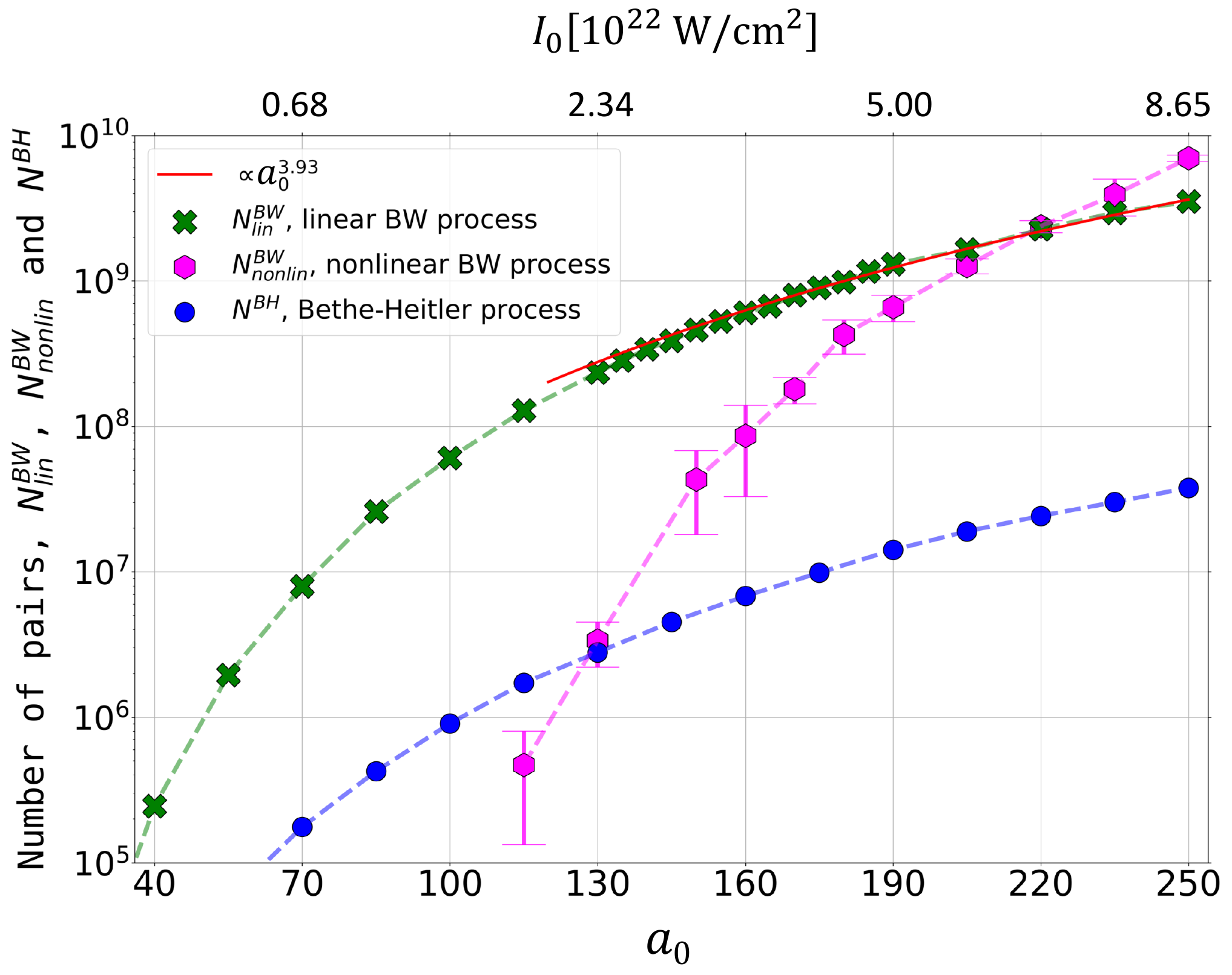}
    \caption{%
        The number of electron-positron pairs created by the linear and nonlinear Breit-Wheeler processes (green crosses and magenta markers, respectively), for the setup shown in \cref{fig:schematic}, at given laser amplitude $a_0$ (and equivalent peak intensity $I_0$).
        Error bars on the nonlinear results indicate statistical uncertainties: see text for details. \tom{The estimated background, electron-positron pairs produced by the Bethe-Heitler process, is shown by blue circles. \bc{The nonlinear Breit-Wheeler pair yield is calculated directly by the PIC code, whereas the linear Breit-Wheeler and Bethe-Heitler pair yields are obtained by post-processing, as described in Methods.} }
    }
    \label{fig:pair_number}
    \end{figure}

The pair yields for the linear and nonlinear processes are compared in \cref{fig:pair_number}. The results for the latter are obtained by performing four simulation runs for each value of $a_0$ with different random seeds: points and error bars give the mean and standard deviation obtained, respectively. At $a_0 < 145$, fewer than ten macropositrons are generated per run, so the corresponding data points are not shown. Our analytical estimates for linear Breit-Wheeler pair creation lead us to expect a yield that scales as $a_0^4$: this is consistent with a power-law fit to the data in \cref{fig:pair_number}, which gives a scaling $\propto a_0^m$, where $m \approx 3.93$. We find that the linear pair yield is significantly larger for $a_0 < 190$.


\tom{The number of positrons produced by the linear Breit-Wheeler process exceeds $10^6$ even for $a_0 = 50$, equivalent to $I_0 = 3.4 \times 10^{21}~\Wcmsqd$, which is well in reach of today's high-power laser facilities.
In order to determine whether this is sufficient to be observed, we estimate the number of pairs produced by the Bethe-Heitler process, which is the principal source of background.
In this process, a $\gamma$ ray with energy $\hbar\omega > 2 mc^2$ creates an electron-positron pair by interacting with the Coulomb field of an atomic nucleus.
The calculation is described in detail in the Methods section.
We sum the pair creation probabilities for each simulated photon, taking into account the distance each photon travels in the plasma channel, to obtain the blue circles in \cref{fig:pair_number}.
The Bethe-Heitler background is smaller than the linear Breit-Wheeler signal by approximately two orders of magnitude, which supports the feasibility of using a plasma channel as a platform for investigating fundamental QED effects.}






 

\section*{Discussion}

\tom{We have shown that} laser-plasma interactions provide a platform to generate and accelerate positrons, created entirely by light and light, at intensities that are within the reach of current high-power laser facilities. While previous research into pair creation at high intensity has focused largely on the nonlinear Breit-Wheeler process, we show that the high density of photons afforded by a laser-plasma interaction \tom{can make} the linear process dominant instead. \tom{As such, the geometry we consider has the potential to enable the first experimental measurement of two-photon pair creation, driven entirely by real photons. More broadly, it motivates reconsideration of the neglect of two-particle interactions in simulations of dense, laser-irradiated plasmas. Such interactions will form a major component of the physics investigated in upcoming high-power laser facilities.} From the theory perspective, our results also motivate investigation of field-driven corrections to the two-photon cross section. The theory for the inverse process, pair annihiliation to two photons, has recently been revisited~\cite{bragin.arxiv.2020}.

\bc{One of our surprising findings, besides the dominance of the linear Breit-Wheeler process, is that the plasma magnetic field preserves its polarity after the two laser pulses collide and pass through each other. The polarity of the magnetic field enables transverse confinement of the positrons within the channel and their acceleration by one of the laser pulses to energies approaching 1~GeV.
We have confirmed this directly for the positrons generated via the nonlinear Breit-Wheeler process.
This should also be the case for the positrons generated via the dominant linear Breit-Wheeler process, because the particles are created inside the channel magnetic field in the presence of a laser pulse, which are the prerequisites for the direct laser acceleration. We therefore expect the positrons to be ejected from the target in the form of collimated jets. The collimation should aid positron detection outside of the target. Moreover, their detection at lower values of $a_0$ should be a clear indicator of the linear Breit-Wheeler process being the source, as the nonlinear process is heavily suppressed for $a_0 \lesssim 150$.}

\bc{Finally, we point out that our observations regarding the dominance of the linear Breit-Wheeler process apply to a range of channel densities. In our simulations, the electron density in the channel is set at $n_{ch} = (a_0/100) 3.8 n_c$, such that it increases linearly with $a_0$ during the intensity scan. Two channel density scans provided in the Supplemental Material show that our observations hold for channel densities that are within a $\pm 20\%$ window of $n_{ch}$.}


\section*{Methods}

\subsection*{Particle-in-cell simulations}

Table~\ref{table:PIC} provides detailed parameters for the simulations presented in the manuscript. Simulations were carried out using the fully relativistic particle-in-cell code EPOCH~\cite{arber2015epoch}. All our simulations are 2D-3V. 

The axis of the structured target is aligned with the axis of the counterpropagating lasers (laser \#1 and laser \#2) at $y = 0$. The target is initialized as a fully-ionized plasma with carbon ions. The bulk electron density is constant during the intensity scan while the electron density in the channel is set at $n_e = (a_0/100)3.8 n_c$. Each laser is focused at the corresponding channel opening. The lasers are linearly polarized with the electric field being in the plane of the simulation. In the absence of the target, the lasers have the same Gaussian profile in the focal spot with the same Gaussian temporal profile. 

We performed additional runs at $a_0 = 190$ with higher spatial resolutions (40 by 40 cells per $\micron$ and 80 by 80 cells per $\micron$). There are no significant variations in the photon spectra for multi-MeV photons and for photons with energies above 50~keV. The electrons that emit energetic photons, as the one whose trajectories in physical and momentum space are shown in \cref{fig:e and p trajectories}(a) and \cref{fig:e and p trajectories}(c), undergo their energy gain without alternating deceleration to non-relativistic energies and re-acceleration. This is likely the reason why they are not subject to a more severe constraint discussed in Refs.~[\citeonline{arefiev.pop.2015,gordon.cpc.2021,tangtartharakul.jcp.2021}] that requires for the cell-size/time-step to be reduced according to the $1/a_0$ scaling in order to achieve convergence.

\begin{table}
\centering
\begin{tabular}{ |p{4cm}|p{4cm}|  }
 \hline
 \multicolumn{2}{|c|}{\textbf{Laser parameters}} \\
 \hline
 Normalized field amplitude & $a_0 = 100 - 190$ \\
 \hline
 Peak intensity range & $I_0 = 1.4 - 4.9\times 10^{22}$ W/cm$^2$ \\
 \hline
 Wavelength & $\lambda_0 = 1$~$\micron$ \\
 \hline
 Focal plane of laser \#1 & $x=- 35$ $\micron$ \\
 \hline
 Focal plane of laser \#2 & $x= + 35$ $\micron$ \\
 \hline
 Laser profile (longitudinal and transverse) & Gaussian \\
 \hline
 Pulse duration (FWHM for intensity)& 50 fs\\
 \hline
 Focal spot size (FWHM for intensity) & 3.6 $\micron$  \\
 \hline
\end{tabular}
\bigskip

\begin{tabular}{ |p{4cm}|p{4cm}|  }
 \hline
 \multicolumn{2}{|c|}{\textbf{Target parameters}} \\
 \hline
 Target thickness (along $y$) & 30 $\micron$ \\
 \hline
 Target length (along $x$) & 70 $\micron$ \\
 \hline
 Channel width & $d_\text{ch} = 5$ $\micron$ \\
 \hline
 Composition & $C^{+6}$ and electrons \\
 \hline
 Channel density & $n_e = 3.8 - 7.1n_{c}$ \\ 
 \hline
 Bulk density & $n_e = 100 n_{c}$ \\
 \hline
\end{tabular}

\bigskip

\begin{tabular}{ |p{4cm}|p{4cm}|  }
 \hline
 \multicolumn{2}{|c|}{\textbf{Other parameters}} \\
 \hline
 Simulation box & 80 $\micron$ in $x$; 36 $\micron$ in $y$\\
 \hline
 Spatial resolution & 40 cells per $\micron$ in $x$\\
 & 20 cells per $\micron$  in $y$\\
 \hline
 Macro-particles per cell & 40 for electrons \\
 & 20 for carbon ions\\
 \hline
\end{tabular}

\caption{2D PIC simulation parameters.}
  \label{table:PIC}
  
\end{table}


\subsection*{Postprocessing algorithm for determination of the linear Breit-Wheeler pair yield}
\label{sec:CollisionAlgorithm}

In order to compute the yield and spatial distribution of the linear Breit-Wheeler pairs, we approximate the photon population as a collection of collimated, monoenergetic beamlets. Discretization into beamlets is achieved by recording the location $(x_0, y_0)$, energy $\epsilon_{\gamma}$, and angle $\theta$ of each photon macroparticle at the time of the emission $t_{emit}$. The photon emission pattern suggests that the emission profile across the channel can be approximated as uniform. We thus represent the emitted photons by a time-dependent distribution function $f = f(x_0, s_{\gamma}, \theta; t_{emit})$, where $s_\gamma \equiv \log_{10}(\epsilon_\gamma/$MeV$)$. It is sufficient to limit our analysis to $-40~\micron \leq x_0 \leq 40~\micron$, $-3 \leq s_\gamma < 3$, and $0^{\circ} \leq \theta \leq 180^{\circ}$. We split each interval into 70 equal segments to obtain $2.6 \times 10^5$ beamlets. We only check for collisions of beamlets propagating to the right with beamlets propagating to the left. The yield is multiplied by a factor of two to account for beamlets with $-180^{\circ} \leq \theta \leq 0^{\circ}$.

The temporal dependence of a beamlet is represented by slices of given density and fixed thickness. For each beamlet pairing, our algorithm finds the interaction volume $V$, the intersections of the beamlet axes and the crossing angle $\psi$.
The pair yield is given by $\Delta N_{\text{lin}}^{\text{BW}} = \sigma_{\gamma \gamma} c ( 1 - \cos \psi ) V \int\! n_1  n_2 dt$, where $n_1$ and $n_2$ are the photon densities in two overlapping slices at the intersection point. In general, the shape of the overlapping region is not rectangular, so the pair creation is visualized by depositing $\Delta N_{lin}^{BW}$ onto a rectangular grid, into cells with centers inside volume $V$. The procedure is repeated for each beamlet pairing to obtain the density of generated pairs.

\begin{figure}[ht]
\centering
\includegraphics[width=0.4\columnwidth]{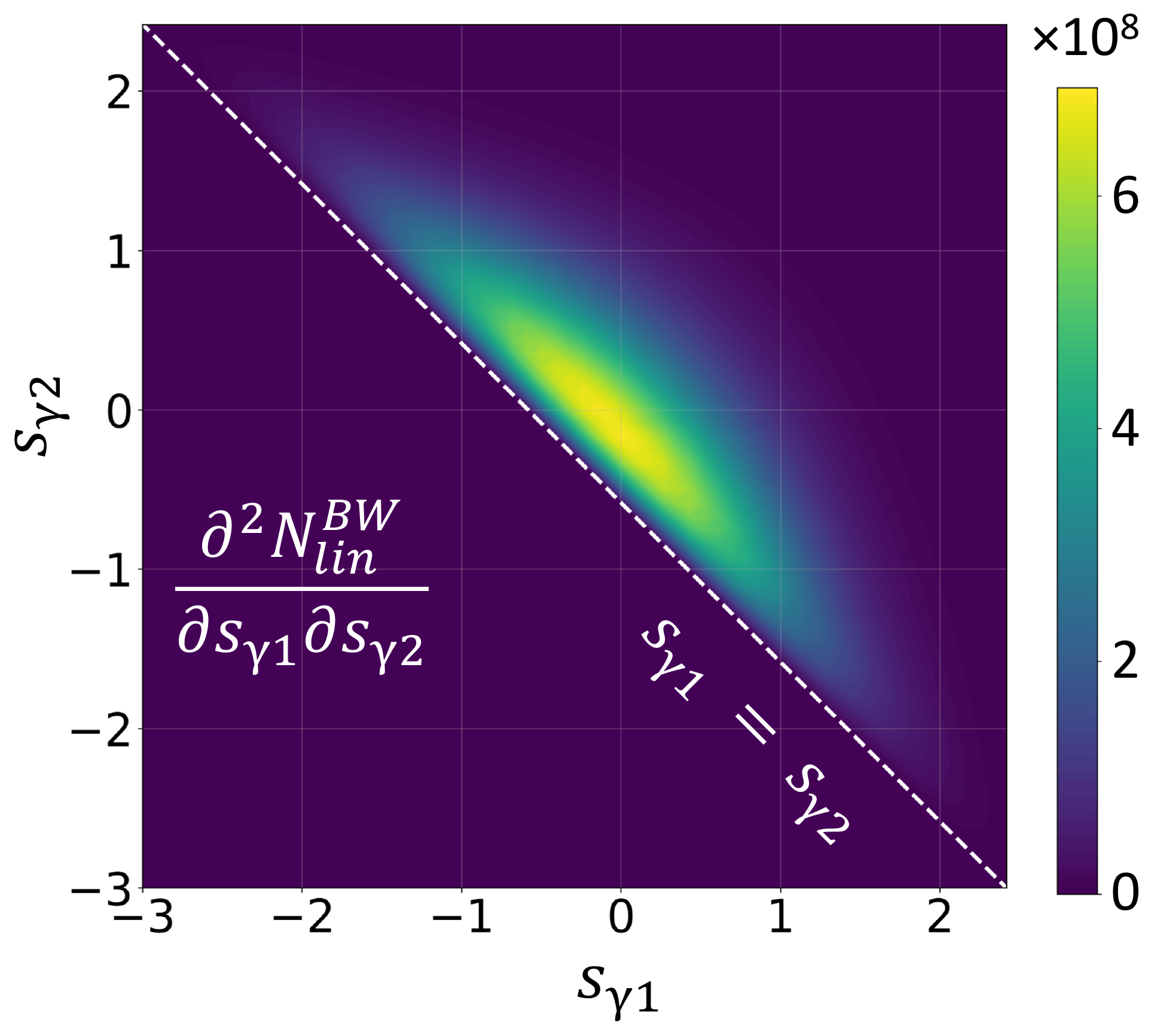}
\caption{Distribution of linear Breit-Wheeler pairs as a function of photon energies for colliding laser pulses with $a_0 = 190$.}
\label{fig:energy vs energy}
\end{figure}

To show that the limitation $-3 \leq s_\gamma < 3$ is justified, we plot the distribution of linear Breit-Wheeler pairs as a function of photon energies. We use $s_{\gamma}$ rather than $\epsilon_{\gamma}$ to capture a wide range of energies. Figure~\ref{fig:energy vs energy} confirms that the pair yield drops off for $|s_{\gamma}| > 2$, which justifies the energy range selected in the manuscript.

The algorithm is a simplification that replaces a direct approach of evaluating all possible collisions of beamlet slices. In a head-on collision, each slice collides with many counterpropagating slices within the interaction volume, which makes the calculation computationally intensive. Our algorithm takes advantage of the fact that the typical duration of beamlet emission, $\tau$, is much longer than the time it takes for photons to travel between the sources emitting the two beamlets, $\ell/c$, where $\ell$ is the distance between the sources in the case of near head-on collision. As shown in the Supplemental Material, our approach is a good approximation as long as $\ell /c \tau < 1$, with the error scaling as $(\ell/c \tau)^2$.

\mc{The postprocessing algorithm neglects the depletion of the photon population due to the linear Breit-Wheeler process. This is justifiable, because only a small fraction of the considered photons actually pair-create (and would therefore be lost). Using the maximum photon density of $n_\gamma \approx 600 n_c$ from \cref{fig:photon density}, we obtain a mean free path with respect to the linear Breit-Wheeler process,
\begin{equation} \label{mfp}
    1/\sigma_{\gamma \gamma} n_\gamma \approx 6 \times 10^4~\micron,
\end{equation}
that is much larger than the characteristic size of the photon cloud of $10~\micron$.
We estimate depletion of the photon population due to the linear Breit-Wheeler process (as a fraction of the initial size) to be smaller than $2 \times 10^{-4}$.}


\subsection*{Estimated background from Bethe-Heitler pair creation}
\label{sec:BetheHeitler}

The principal source of background in a prospective measurement of linear (or nonlinear) Breit-Wheeler pair creation is the Bethe-Heitler process, wherein a photon with energy $\hbar\omega > 2 m c^2$ produces an electron-positron pair on collision with an atomic nucleus~\cite{bethe.prsa.1934}.
\tom{In order to estimate the contribution from this process, we sum the pair creation probability for each macrophoton in the simulation: $N_\text{BH} = \sum_k w_k P_{k,\text{BH}}$, where $w_k$ is the weight of the $k$th photon, scaled assuming that the third dimension has size $5~\micron$.
The probability $P_{k,\text{BH}} = n_i \ell_k \sigma_\text{BH}$, where $n_i$ is the density of carbon ions in the channel and $\ell_k$ is the distance the photon travels before it leaves the channel.
We estimate $n_i$ and $\ell_k$ using the unperturbed properties of the channel, i.e. those at the start of the simulation, and taking into account the photon's point of emission and direction of propagation.
Thus $n_i = 3.8 a_0 n_c / (100 Z)$.}
We approximate the cross section $\sigma_\text{BH}$ by that for an unscreened, fully ionized, point carbon nucleus (formula 3D-0000 in Ref.~[\citeonline{motz.rmp.1969}] with $Z = 6$).
The functional dependence of the cross section on the normalized photon energy $\gamma = \hbar\omega / (m c^2)$ is given by $\sigma_\text{BH}(\gamma) \simeq \alpha r_e^2 Z^2 (2\pi/3) [(\gamma - 2)/ \gamma]^3$ for $\gamma - 2 \ll 1$ and $\sigma_\text{BH}(\gamma) \simeq \alpha r_e^2 Z^2 [28 \ln(2\gamma) / 9 - 218/27]$ for $\gamma \gg 1$, where $r_e$ is the classical electron radius~\cite{motz.rmp.1969}.
Our results are shown as blue circles in \cref{fig:pair_number}.
\tom{This estimate neglects contributions from pair creation in the plasma bulk, which can be controlled by reducing the thickness of the channel walls.
Furthermore, the difference in magnitude between background and signal is sufficiently large that it provides a margin of safety.}




\section*{Supplemental material}

\subsection*{Channel density scan}

\begin{table}[htb]
\centering
\caption{Number of pairs at $a_0=160$ for different electron densities, $n_e$, in the channel. The density $n_e$ is given in terms of $n_{ch} = (a_0/100) 3.8 n_c$.}
\label{tab:160}
\begin{tabular}{llll}
\hline
\multicolumn{1}{|c|}{$n_e / n_{ch}$} & \multicolumn{1}{c|}{0.8} & \multicolumn{1}{c|}{1.0} & \multicolumn{1}{c|}{1.2} \\ \hline \hline
\multicolumn{1}{|l|}{Linear BW pairs ($\times 10^8$)} & \multicolumn{1}{l|}{$5.89$} & \multicolumn{1}{l|}{$6.05$} & \multicolumn{1}{l|}{$6.21$} \\ \hline
\multicolumn{1}{|l|}{Nonlinear BW pairs ($\times 10^8$)} &  \multicolumn{1}{l|}{$1.06$} & \multicolumn{1}{l|}{$0.86$} & \multicolumn{1}{l|}{$0.38$} \\ \hline        
\end{tabular}
\end{table}

\bc{In our simulations, the electron density in the channel is set at $n_{ch} = (a_0/100) 3.8 n_c$. This value was chosen to achieve prolonged laser propagation inside the channel, such that each laser pulse can generate ultrarelativistic electrons without becoming significantly depleted prior to the collision. In order to show that the observed trend is valid for a range of channel densities, we have performed channel density scans for $a_0 = 160$ and $a_0 = 190$. The pair yield for the linear and nonlinear Breit-Wheeler processes is given in \cref{tab:160} for $a_0 = 160$ and in \cref{tab:190} for $a_0 = 190$. The ratio of the linear to nonlinear pair yield increases for $0.8 \leq n_e/n_{ch} \leq 1.2$ as we reduce $a_0$ from 190 to 160. This is the trend that is reported in the main text, which indicates that our observations apply to a range of channel densities and no fine-tuning is necessary.}

\begin{table}[htb]
\centering
\caption{Number of pairs at $a_0=190$ for different electron densities, $n_e$, in the channel. The density $n_e$ is given in terms of $n_{ch} = (a_0/100) 3.8 n_c$.}
\label{tab:190}
\begin{tabular}{llll}
\hline
\multicolumn{1}{|c|}{$n_e / n_{ch}$}  & \multicolumn{1}{c|}{0.8} & \multicolumn{1}{c|}{1.0} & \multicolumn{1}{c|}{1.2} \\ \hline \hline
\multicolumn{1}{|l|}{Linear BW pairs ($\times 10^8$)} &  \multicolumn{1}{l|}{$17.0$} & \multicolumn{1}{l|}{$13.1$} & \multicolumn{1}{l|}{$11.8$} \\ \hline
\multicolumn{1}{|l|}{Nonlinear BW pairs ($\times 10^8$)} &  \multicolumn{1}{l|}{$9.2$} & \multicolumn{1}{l|}{$6.6$} & \multicolumn{1}{l|}{$4.4$} \\ \hline    
\end{tabular}
\end{table}

\subsection*{Distribution of nonlinear Breit-Wheeler pairs}

\begin{figure}[htb]
\centering
\includegraphics[width=0.5\columnwidth]{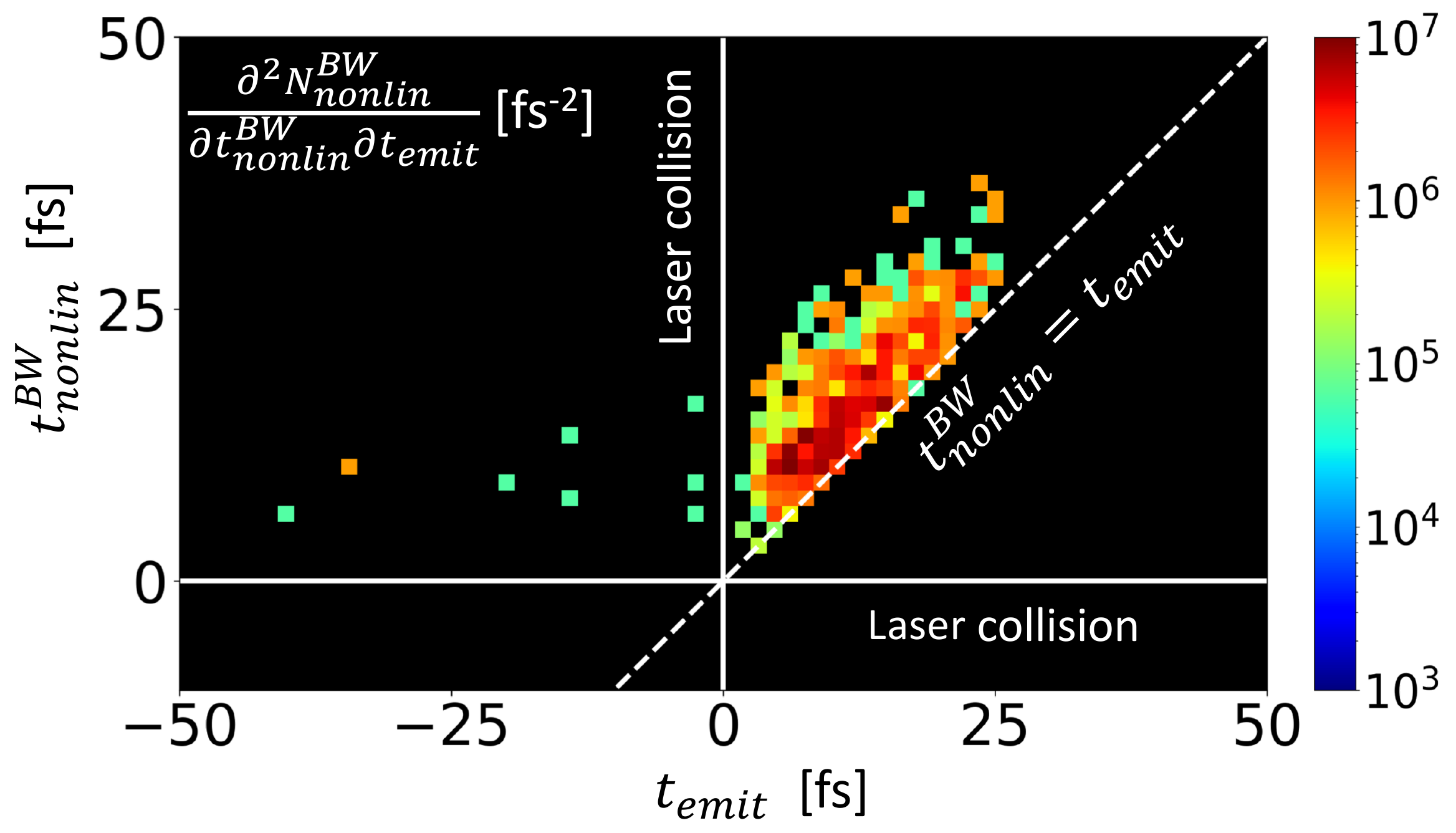}
\caption{Distribution function $\partial ^2 N^{BW}_{nonlin}/(\partial t^{BW}_{nonlin} \partial t_{emit})$ [fs$^{-2}$] on the number of electron-positron pairs over the time $t^{BW}_{nonlin}$ of pair creation and the $t_{emit}$ of the parent photons. The peak amplitude of each laser peak is $a_0 = 190$.}
\label{fig:time_time_nonlinear}
\end{figure}

Figure~\ref{fig:time_time_nonlinear} provides additional information on the production of nonlinear Breit-Wheeler pairs. The horizontal scale shows the emission time, $t_{emit}$, of energetic photons that go on to produce pairs by interacting with the laser photons. The energetic photons are generated and subsequently propagated as particles by the PIC code. The time of the pair production, $t_{nonlin}^{BW}$, is shown along the vertical scale. The number of pairs for each pairing of $t_{emit}$ and $t_{nonlin}^{BW}$ is color-coded. The pair production is directly computed by the PIC code. The numbers shown in the figure are obtained by assuming that the spatial scale along the $z$ axis is equal to the initial width of the channel. 

The vast majority of the nonlinear pairs are produced by photons emitted after the two lasers collide. Indeed, $t_{emit} = 0$ is the time when the two laser beams begin to collide. Most of the pairs are located at $t_{emit} > 0$, which indicates that the parent photons were generated after the collision.


\subsection*{Strong-field modifications to the linear Breit-Wheeler cross section}


The cross section we use to determine the number of electron-positron pairs created by the linear Breit-Wheeler process, Eq.~(1), is calculated assuming that the photons are in vacuum. However, our calculations show that the pairs are created within the plasma channel as the laser pulses overlap, where the background electromagnetic field is strong. The presence of a strong magnetic field~\cite{ng.prd.1977,kozlenkov.jetp.1986} or plane EM wave is known to modify the cross section for this process. Calculations of the cross section in the latter case have focused on changes at moderate $a_0$~\cite{denisenko.lp.2008} or on resonance features~\cite{hartin.ijmpa.2018,hartin.2006}. Resonances occur where the intermediate fermion is on mass shell, which corresponds to the incoherent combination of nonlinear pair creation, followed by photon absorption by one of the daughter fermions, in the high-intensity regime. This contribution, which is effectively driven by a single photon, is already counted as our simulations include nonlinear pair creation; photon absorption by an electron in a strong field~\cite{ritus.jslr.1985,ilderton.prd.2019} is suppressed unless the photon and electron are aligned within electron's emission cone. (See supporting simulations in Ref.~[\citeonline{blackburn.arxiv.2020}].)

As such, in order to estimate the effect of the strong field at $a_0 \gg 1$, we use the cross section for two-photon pair creation in a constant, crossed field given in Sec. 5.7 of Ref.~[\citeonline{bks}]. If two-photon pair creation is kinematically allowed, i.e. $\varsigma > 1$, corrections to the cross section scale as $(\chi_\gamma / \varsigma)^2$, where $\chi_\gamma$ is the quantum nonlinearity parameter~\cite{bks}: in the scenario under consideration here, the photons which undergo linear pair creation have MeV energies and $\chi_\gamma \ll 1$, and therefore these corrections can be neglected. On the other hand, the fact the laser-plasma interactions provide a platform for prolific two-photon pair creation in a region of strong EM field, as our results show, motivates a more general treatment that can investigate where these field-driven corrections become substantial. We note that a related process, pair annihiliation to two photons in a pulsed, plane EM wave, has recently been revisited~\cite{bragin.arxiv.2020}.


\subsection*{Pair production algorithm for head-on collisions of beamlets }

Our algorithm for computing the pair production yield due to the linear Breit-Wheeler process leverages the fact that the colliding photons are emitted over an extended period of time compared to the characteristic travel time between the emission locations. In what follows, we illustrate its implementation for head-on collisions of two beamlets. Near head-on collisions are the biggest contributor to the pair yield and this is also the regime that greatly benefits from our simplified approach in terms of computational efficiency.

We are considering a head-on collision of two counterpropagating beamlets that have the same transverse area $S$. The first beamlet is being emitted at $x = x_1$ and it propagates in the positive direction. It is convenient to use the emission time $\tau$ as a marker for the photons in each beamlet. The corresponding photon density in the first beamlet is then $n_1 (\tau_1)$. The counterpropagating beamlet is emitted at $x = x_2 > x_1$ and its photon density is $n_2 (\tau_2)$. The total pair yield by these two beamlets interacting with each other is
\begin{equation} \label{number of pairs -- direct}
   \Delta  N_{pairs} = \sigma_{\gamma \gamma} S c^2\int_{- \infty}^{+ \infty} d \tau_1 n_1 ( \tau_1 ) \left[ \int_{\tau_1 - l/c}^{\tau_1 + l/c} d \tau_2 n_2 (\tau_2) \right],
\end{equation}
where
\begin{equation}
    l \equiv x_2 - x_1.
\end{equation}
Equation~(\ref{number of pairs -- direct}) presents a direct approach to calculating the pair yield and it involves a double integral.

In our case, the typical beamlet emission lasts longer than $l/c$, which suggests a possible simplification of replacing $n_2 (\tau_2)$ with $n_2 (\tau_1)$. The inner integral in Eq.~(\ref{number of pairs -- direct}) can then be directly evaluated and we find that
\begin{equation} \label{number of pairs -- av}
    \Delta N_{pairs} \approx 2c \sigma_{\gamma \gamma} V \int_{- \infty}^{+ \infty} n_1 ( \tau ) n_2 ( \tau ) d \tau  ,
\end{equation}
where $V \equiv l S$ is the interaction volume. Note that the two beamlets are interchangeable in this expression. In order to estimate the error, we expand $n_2 (\tau_2)$ around $\tau_2 = \tau_1$ and retain linear and quadratic terms in the expansion. The time integral of the linear term in Eq.~(\ref{number of pairs -- direct}) is equal to zero, which means that the error in our approach is determined by the quadratic term. We thus estimate that the relative error in the number of produced pairs scales as $(l/c \Delta \tau)^2$, where $\Delta \tau$ is the characteristic duration of beamlet emission.

We compute the spatial distribution of pairs by simply assigning the pair density
\begin{equation} \label{n_pairs}
    \Delta n_{pairs} = \Delta N_{pairs} / V 
\end{equation}
to each position within the interaction volume. We call this the \emph{average density method}. A direct approach would however require us to compute the following integral at each position along the $x$ axis:
\begin{equation} \label{direct}
    \Delta n^{direct}_{pairs} (x) = 2 c \sigma_{\gamma \gamma} \int_{-\infty}^{+\infty} n_1 \left( t_1 \right) n_2 \left(t_2 \right) dt,
\end{equation}
where
\begin{eqnarray}
&& t_1 = t - (x - x_1)/c, \\
&& t_2 = t - (x_2 - x)/c.
\end{eqnarray}
The direct approach is much more demanding computationally.

\begin{figure}[htb]
\centering
\includegraphics[width=0.5\columnwidth]{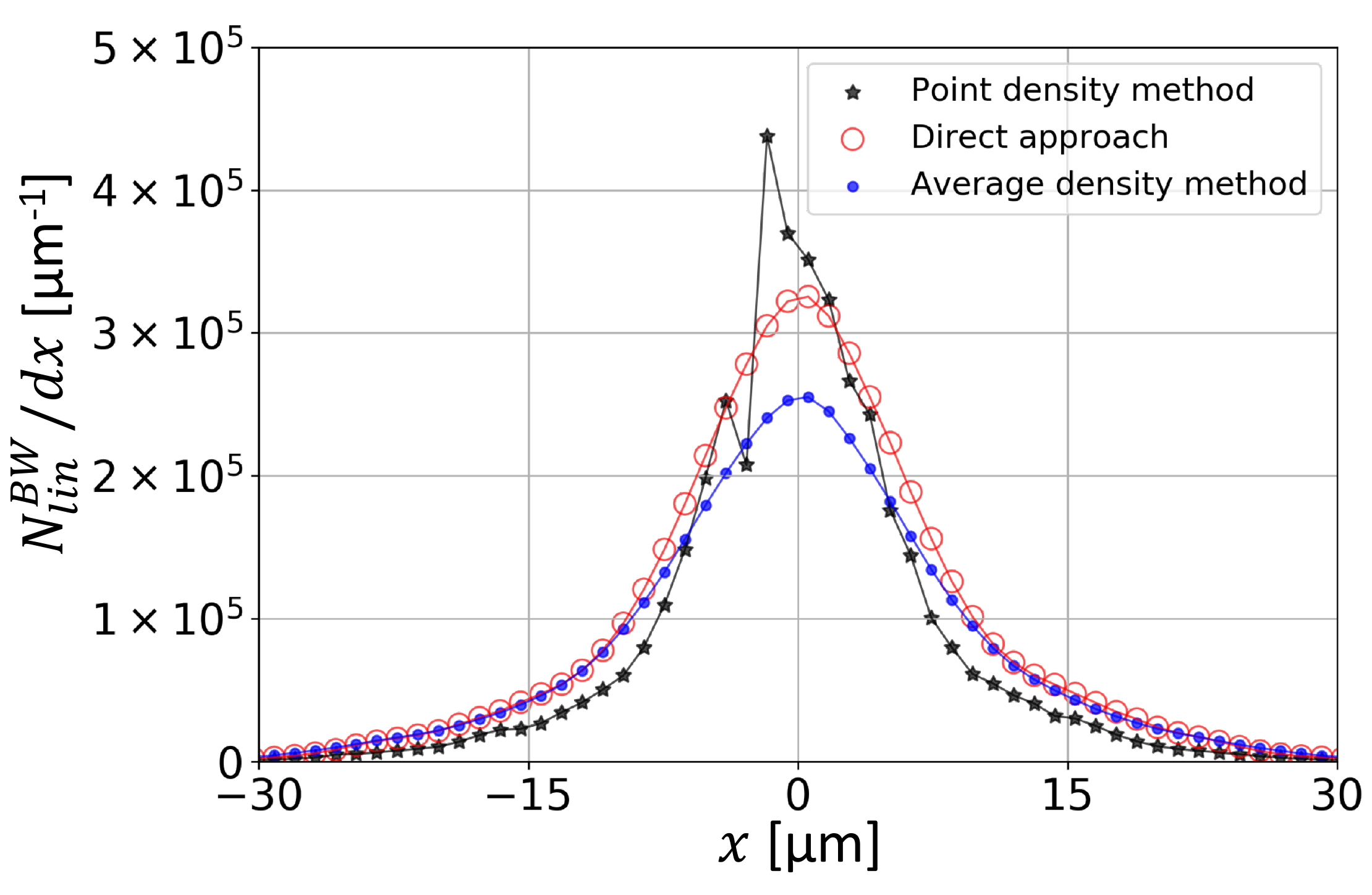}
\caption{Spatial distribution of linear Breit-Wheeler pairs produced in a head-on collision of spatially distributed beamlets. The three curves represent three different approaches: direct approach (open circles), average density method (solid circles), and point density method (star markers).}
\label{fig:benchmark}
\end{figure}

Even though our approach for calculating the spatial distribution of pairs is deliberately crude for a single pair of beamlets, it is effective when applied to a large ensemble of spatially distributed beamlets. As an example, we have carried out calculations for approximately 3000 spatially distributed beamlets, where 1573 beamlets are directed to the right and 1590 beamlets are directed to the left. The photon energy range is $0.19 \mbox{ MeV} < \epsilon_{\gamma} < 2.05 \mbox{ MeV}$. We used our PIC simulation to generate these beamlets by selecting photons emitted with $|\theta| < 5.15^{\circ}$ or $|\theta - \pi| < 5.15^{\circ}$. The average density method that uses $\Delta n_{pairs}$ from Eq.~(\ref{n_pairs}) for each beamlet-beamlet collision gives the curve shown with small solid circles. The direct approach detailed by Eq.~(\ref{direct}) gives the curve shown with open circles. The two curves have a similar shape, but the direct approach took almost two orders of magnitude longer in terms of computational time. The relative difference in the total number of pairs between the two methods is less than 17\%. 

Our method evenly distributes the generated pairs over the interaction volume, which is the key to achieving a good agreement with the direct but more computationally expensive approach. In order to illustrate this aspect, we performed another calculation. In this case, all of the pairs produced by two beamlets are placed into the center of the interaction volume without being evenly distributed. We call this the \emph{point density method}. The density is calculated as $\Delta N_{pairs} / S \Delta x$, where $\Delta N_{pairs}$ is given by Eq.~(\ref{number of pairs -- av}) and $\Delta x$ is the thickness of slices that we use for spatial discretization into beamlets. The result of this procedure is shown with star markers in Fig.~\ref{fig:benchmark}. There are no significant savings in terms of computational costs compared to the average density method. The characteristic width of the spatial distribution shows considerable deviation from that for the direct approach.

\bibliography{sample}



\section*{Acknowledgements}

This research was supported by AFOSR (Grant No. FA9550-17-1-0382). Simulations were performed with EPOCH (developed under UK EPSRC Grants EP/G054950/1, EP/G056803/1, EP/G055165/1 and EP/ M022463/1) using high performance computing resources provided by TACC.

\section*{Author contributions statement}

\ava{Y.H. performed the PIC simulations and developed the algorithm to calculate the yield of linear Breit-Wheeler pairs. A.A, T.B., and T.T. conceived the original idea. All authors contributed to writing and reviewing the manuscript.}

\section*{Data availability}

The datasets generated during and/or analyzed during the current study are available from the corresponding author on reasonable request.

\section*{Code availability}

PIC simulations were performed with an open-source, open-acces PIC code EPOCH~\cite{arber2015epoch}. The photon-photon collision code is based on the algorithm detailed in the Methods and Supplemental material sections. It is available from the corresponding author on reasonable request.




\end{document}